\documentclass[aps,prl,twocolumn,floats,epsfig,showpacs]{revtex4-1}

\usepackage{bbm}
\usepackage{amssymb}
\usepackage{ifpdf}
\usepackage{color}
\usepackage{graphicx}
\usepackage{amsmath}
\usepackage{amsfonts}
\usepackage{hyperref}
\usepackage{dcolumn}
\usepackage{bm}
\usepackage{xcolor}

\newcommand{\Z}{{\mathbb{Z}}}

\newcommand{\1}{{\mathbbm{1}}}

\definecolor{red}{rgb}{0.7,0,0}
\definecolor{green}{rgb}{0.,0.35,0.}
\definecolor{blue}{rgb}{0.2,0.2,0.7} 
\definecolor{black}{rgb}{0.15,0.15,.15}

\begin{document}

\title{Atomic Quantum Simulation of Dynamical Gauge Fields coupled to Fermionic Matter: \\ From String Breaking to Evolution after a Quench}
\author{D.\ Banerjee$^1$, M.\ Dalmonte$^{2,3}$, M.\ M\"uller$^4$, E.\ Rico$^{2,3}$, P.\ Stebler$^1$, U.-J.\ Wiese$^1$, and P.\ Zoller$^{2,3,5}$}
\affiliation{$^1$Albert Einstein Center, Institute for Theoretical Physics, Bern University, CH-3012, Bern, Switzerland \\
$^2$Institute for Quantum Optics and Quantum Information of the Austrian Academy of Sciences, A-6020 Innsbruck, Austria \\
$^3$ Institute for Theoretical Physics, Innsbruck University, A-6020 Innsbruck, Austria\\
$^4$Departamento de Fisica Teorica I, Universidad Complutense, 28040 Madrid, Spain \\
$^5$ Joint Quantum Institute:   National Institute of Standards and Technology,
and University of Maryland, College Park, Maryland 20742, USA}

\begin{abstract}
Using a Fermi-Bose mixture of ultra-cold atoms in an optical lattice, we construct a quantum simulator for a $U(1)$ gauge theory coupled to fermionic matter. The construction is based on quantum links which realize continuous gauge symmetry with discrete quantum variables. At low energies, quantum link models with staggered fermions emerge from a Hubbard-type model which can be quantum simulated. This allows us to investigate string breaking as well as the real-time evolution after a quench in gauge theories, which are inaccessible to classical simulation methods.
\end{abstract}

\maketitle

Recently, the condensed matter and atomic physics communities have mutually benefited from synergies emerging from the quantum simulation of strongly correlated systems using atomic setups \cite{lewenstein2012ultracold,Cirac2012a,Bloch2012,Blatt2012}. In particular, physically interesting quantum many-body systems, which can not be solved with classical simulation methods, are becoming accessible to analog or digital quantum simulation with cold atoms, molecules, and ions. In the future, quantum simulators may also enable us to address currently unsolvable problems in particle physics, including the real-time evolution of the hot quark-gluon plasma emerging from a heavy-ion collision or the deep interior of neutron stars \cite{Rajagopal2000}.

The challenge on the atomic physics side is to find a physical implementation of gauge theories with cold atoms, and to identify possible atomic setups representing dynamical gauge fields coupled to fermionic matter. Below we provide a toolbox for a $U(1)$ lattice gauge theory (LGT) using atoms in optical lattices \cite{Bloch2012,lewenstein2012ultracold}. Here fermionic atoms represent matter fields. They hop between lattice sites and interact with dynamical gauge fields on the links embodied by bosonic atoms. The LGT to be implemented is a so-called quantum link model (QLM)\cite{Horn1981a,Orland1990a,Chandrasekharan1997a}, where the fundamental gauge variables are represented by quantum spins. QLMs extend the concept of Wilson's LGT \cite{Wilson1974}. In particle physics they provide an alternative non-perturbative formulation of dynamical Abelian and non-Abelian gauge field theories \cite{Chandrasekharan1997a,Brower1999,Brower2004a}. QLMs are also relevant in condensed matter contexts, like spin liquids and frustrated systems \cite{Duan2003,Hermele2004,Levin2005}. Their Hamiltonian formulation provides a natural starting point for quantum simulation protocols based on atomic gases in optical lattices \cite{Buchler2005,Tewari2006,weimer:2010,Cirac2010a,Kapit2011}. We will illustrate atomic quantum simulation of an Abelian QLM in a 1D setup, demonstrating both dynamical string breaking and the real-time evolution after a quench, which are also relevant in QCD. The quantum simulator discussed below makes the corresponding real-time dynamics, which is exponentially hard for classical simulations based on Wilson's paradigm \cite{Troyer2005}, accessible to atomic experiments.

Cold quantum gases provide a unique experimental platform to study many-body dynamics of isolated quantum systems. In particular, cold atoms in optical lattices realize Hubbard dynamics for both bosonic and fermionic particles, where the single particle and interaction terms can be engineered by external fields. The remarkable experimental progress is documented by the quantitative determination of phase diagrams in strongly interacting regimes, the study of quantum phase transitions, and non-equilibrium quench dynamics \cite{Nascimbene2010,Trotzky2010,van2012feynman,Zhang2012,Trotzky2012a}. One of the most exciting recent developments are 
\emph{synthetic gauge fields with atoms}, which promises the realization of strongly correlated many-body phases, such as, e.g., the fractional quantum Hall effect with atoms \cite{jaksch2003creation,Goldman2009,Lin2011,Dalibard2011,Aidelsburger2011,cooper2011}. A fermion that is annihilated by $\psi_y$ and recreated by $\psi^\dagger_x$ at a neighboring site $x$, which propagates in the background of a classical Abelian vector potential $\vec A$ gives rise to the hopping term $\psi_x^\dagger u_{xy} \psi_y$ with $u_{xy} = \exp(i \varphi_{xy})$. Hopping between the adjacent lattice sites $x$ and $y$ accumulates the phase $\varphi_{xy} = \int_x^y d\vec l \cdot \vec A$. The hopping term is invariant against $U(1)$ gauge transformations $\vec A^\prime = \vec A - \vec \nabla \alpha$\cite{Cohen1977,supmat}. When a fermion hops around a lattice plaquette $\langle w x y z \rangle$, it picks up a gauge invariant magnetic flux phase $\exp(i \Phi) = u_{wx} u_{xy} u_{yz} u_{zw}$, with $\Phi = \int d^2 \vec f \cdot \vec \nabla \times \vec A$. We emphasize that these synthetic gauge fields are \emph{$c$-numbers} mimicking an external magnetic field for the (neutral) atoms.

Instead, here we are interested in \emph{dynamical} gauge fields as they arise in particle physics \cite{Kogut1983}. The corresponding fundamental bosonic degrees of freedom $U_{xy}$ are no longer related to an underlying classical background field $\vec A$, but represent quantum operators associated with the lattice links. The hopping of the fermions is now mediated by the bosonic gauge field via the term $\psi_x^\dagger U_{xy} \psi_y$, which is invariant under local changes of matter and gauge degrees of freedom 
$U_{xy}^\prime = V^\dagger U_{xy} V = \exp(i \alpha_x) U_{xy} \exp(-i \alpha_y)$, 
$\psi_x^\prime = V^\dagger \psi_x V = \exp(i \alpha_x) \psi_x, \quad V =
\prod_x \exp\left(i \alpha_x G_x \right)$,
and $G_x = \psi_x^\dagger \psi_x - \sum_i \left(E_{x,x+\hat i} - E_{x-\hat i,x}\right)$.
Here $E_{x,x+\hat i}$ is an electric field operator associated with the link connecting $x$ and $y = x + \hat i$, where $\hat i$ is a unit-vector in the $i$-direction. $G_x$ is the generator of gauge transformations (see \cite{supmat} for a detailed discussion). Gauge invariant physical states must obey Gauss' law, $G_x|\Psi\rangle = 0$, which is the lattice variant of $\vec \nabla \cdot \vec E = \rho = \psi^\dagger \psi$. To ensure gauge covariance of $U_{xy}$, it must obey 
$[E_{xy},U_{xy}] = U_{xy}$. The Hamiltonian representing the electric and magnetic field energy of a compact $U(1)$ LGT, $H=\frac{g^2}{2} \! \sum_{\langle x y \rangle} E_{xy}^2 - \frac{1}{4 g^2} \sum_{\langle w x y z \rangle} \:\left(U_{wx} U_{xy} U_{yz}
U_{zw} + \mathrm{h.c.}\right)$,
is gauge invariant, i.e.\ $[H,G_x] = 0$. In Wilson's LGT, the link variables $U_{xy} = \exp(i \varphi_{xy}) \in U(1)$ are still complex phases, and $E_{xy} = - i \partial/\partial \varphi_{xy}$. Since $U_{xy}$ is a continuous variable, which implies an infinite-dimensional Hilbert space per link, it is not clear how to implement it in ultra-cold matter, where one usually deals with discrete degrees of freedom in a finite-dimensional Hilbert space.

{\em Quantum link models} offer an attractive framework for the quantum simulation of dynamical gauge fields \cite{Chandrasekharan1997a,Brower1999,Brower2004a}. They extend the concept of a LGT to systems of discrete quantum degrees of freedom with only a finite-dimensional Hilbert space per link. In contrast to the Wilson formulation,  QLMs resemble a quantum rather than a classical
statistical mechanics problem. The relation $[E_{xy},U_{xy}] = U_{xy}$ is then realized by a quantum link operator $U_{xy} = S_{xy}^+$ which is a raising operator for the electric flux $E_{xy} = S_{xy}^3$ associated with the link connecting neighboring lattice sites $x$ and $y$. A local $SU(2)$ algebra is generated by a quantum spin $\vec S_{xy}$ with just $2S+1$ states per link~\cite{supmat}. We will consider quantum links with $S = \frac{1}{2}$ or 1. In the classical limit $S \rightarrow \infty$  QLMs reduce to the Hamiltonian formulation \cite{Kogut1975,Banks1976} of Wilson's LGT.

\begin{figure}[tbp]
\includegraphics[width=0.8\columnwidth]{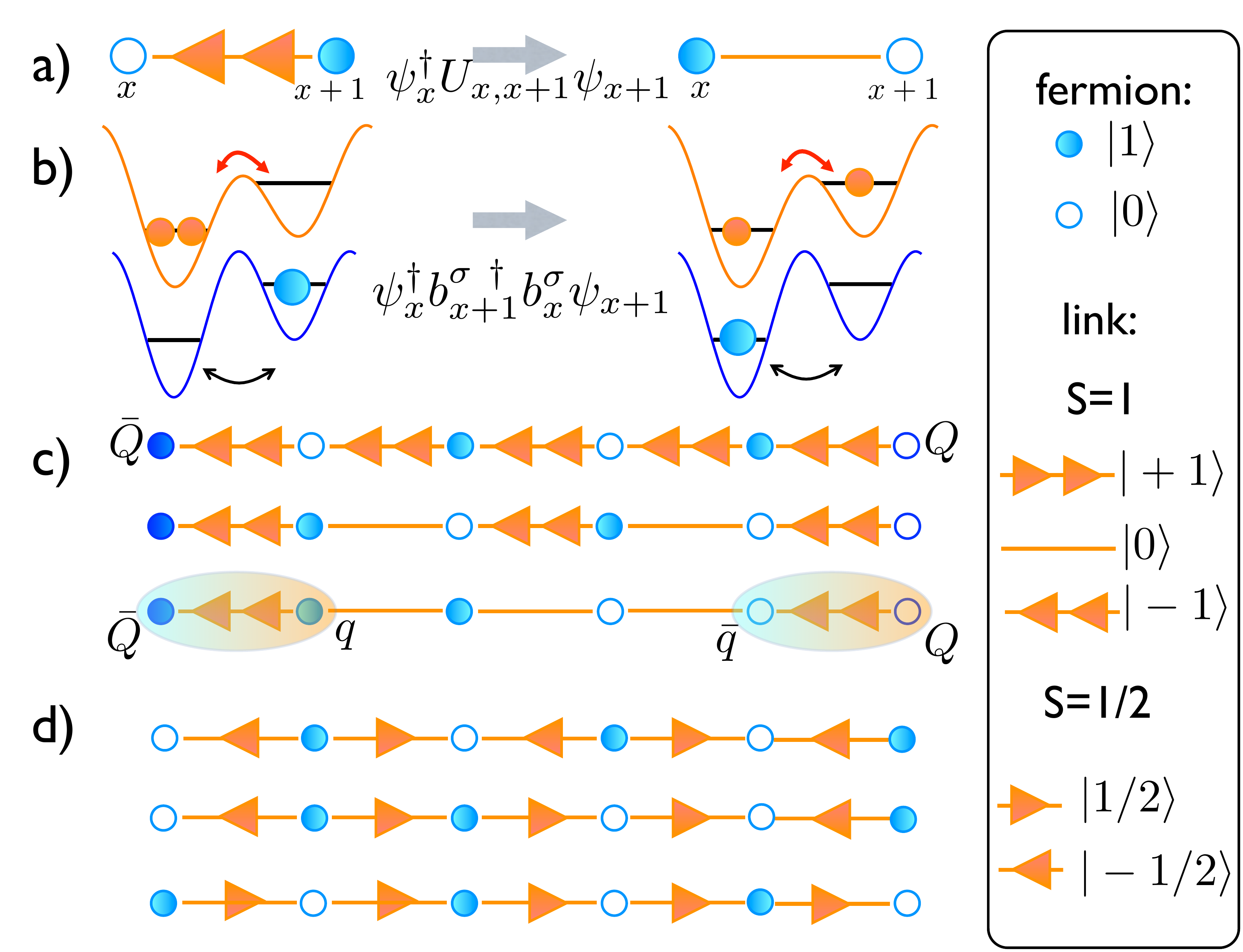}
\caption{[Color online] { a) Correlated hop of a fermion assisted by $U_{x,x+1}\equiv S^{+}_{x,x+1}$  consistent with Gauss' law in a QLM with spin $S = 1$. b) Realization of the process in a) with bosonic and fermionic atoms in an optical super-lattice (see text). c) Breaking of a string connecting a static $\bar Q Q$ pair: from an unbroken string (top), via fermion hopping (middle), to two mesons separated by vacuum (bottom). d) From a parity-invariant staggered flux state (top), via fermion hopping (middle), to the vacuum with spontaneous parity breaking.}}
\label{cartoon}
\end{figure}
{\em The implementation of quantum link models} in ultra-cold matter requires the realization of a gauge invariant Hamiltonian accompanied by the corresponding Gauss law. Here, we present a general procedure to obtain $U(1)$ QLMs including both gauge and matter fields. To illustrate our method, we focus on a simple example, a 1D $U(1)$ QLM coupled to so-called staggered fermions with the Hamiltonian 
\begin{eqnarray}
\label{Hamilton}
H&=&- t \sum_x \left[\psi_x^\dagger U_{x,x+1} \psi_{x+1} + \mathrm{h.c.}\right]
\nonumber \\
&+&m \sum_x (-1)^x \psi_x^\dagger \psi_x + \frac{g^2}{2} \sum_x E_{x,x+1}^2.
\end{eqnarray}
Here $t$ is the hopping pa\-ra\-me\-ter (see Fig.~\ref{cartoon}a), $m$ is the fermion mass, and $g$ is the gauge coupling. In this case, the gauge generator is given by $\widetilde G_x = G_x + \frac{1}{2}\left[(-1)^x - 1 \right]$. Staggered fermions are analogous to spinless fermions at half-filling in condensed matter physics. The corresponding vacuum represents a filled Dirac sea of negative energy states. For $S = 1$, $t = 0$, and $m > 0$ the vacuum state has $E_{x,x+1} = 0$ and $\psi_x^\dagger \psi_x = \frac{1}{2}\left[1 - (-1)^x\right]$. The corresponding vacuum energy of a system with $L$ sites is $E_0 = - m L/2$. The above Hamiltonian resembles the Schwinger model \cite{Coleman1976}. For $S = 1$ it shares the non-perturbative phenomenon of string breaking by dynamical $q \bar q$ pair creation with QCD \cite{pepe2009}. An external static quark-anti-quark pair $\bar Q Q$ (with the Gauss law appropriately taken into account) is connected by a confining electric flux string (Fig. \ref{cartoon}c, top), which manifests itself by a large value of the electric flux. For $t = 0$, the energy of this state is $E_{\text{string}} - E_0 = g^2 (L - 1)/2$, and the flux is given by
$\langle \sum_x E_{x,x+1}\rangle = - L +1$. At sufficiently large $L$, the string's potential energy is converted into kinetic energy by fermion hopping, which amounts to the creation of a dynamical quark-anti-quark pair $q \bar q$. 
In this process, which is known as string breaking, an external static anti-quark $\bar Q$ pairs up with a dynamical quark to form a $\bar Q q$ meson. For $t = 0$, the resulting two-meson state of 
Fig. \ref{cartoon}c (bottom) has an energy $E_{\text{mesons}} - E_0 = g^2 + 2m$ and a small flux $\langle \sum_x E_{x,x+1} \rangle = - 2$. The energy difference $E_{\text{string}} - E_{\text{mesons}} = g^2 (L - 3)/2 - 2m = 0$ determines the length $L = 4m/g^2 + 3$ at which the string breaks.

Another non-perturbative process of interest in particle physics is the real-time evolution after a quench. In particular, the quark-gluon plasma created in a heavy-ion collision quickly returns to the ordinary hadronic vacuum. This is accompanied by the spontaneous breakdown of the quark's chiral symmetry. The dynamics after a quench can be quantum simulated by using the $S = \frac{1}{2}$ representation for the electric flux (which mimics the Schwinger model at vacuum angle $\theta = \pi$ \cite{Coleman1976}). In that case, like chiral symmetry in QCD, for $m > 0$ parity is spontaneously broken, at least for small $t$, for more details see \cite{supmat}. A quenched parity-invariant staggered flux state, which evolves into the true vacuum with spontaneous parity breaking, is schematically illustrated in Fig.~\ref{cartoon}d. In this case, the electric flux represents an order pa\-ra\-me\-ter for spontaneous parity breaking, which is expected to perform coherent oscillations. This is similar to the time evolution after a quench starting from a disoriented chiral condensate in QCD \cite{Rajagopal1993}.

\begin{figure}[tbp]
\includegraphics[width=0.34\textwidth]{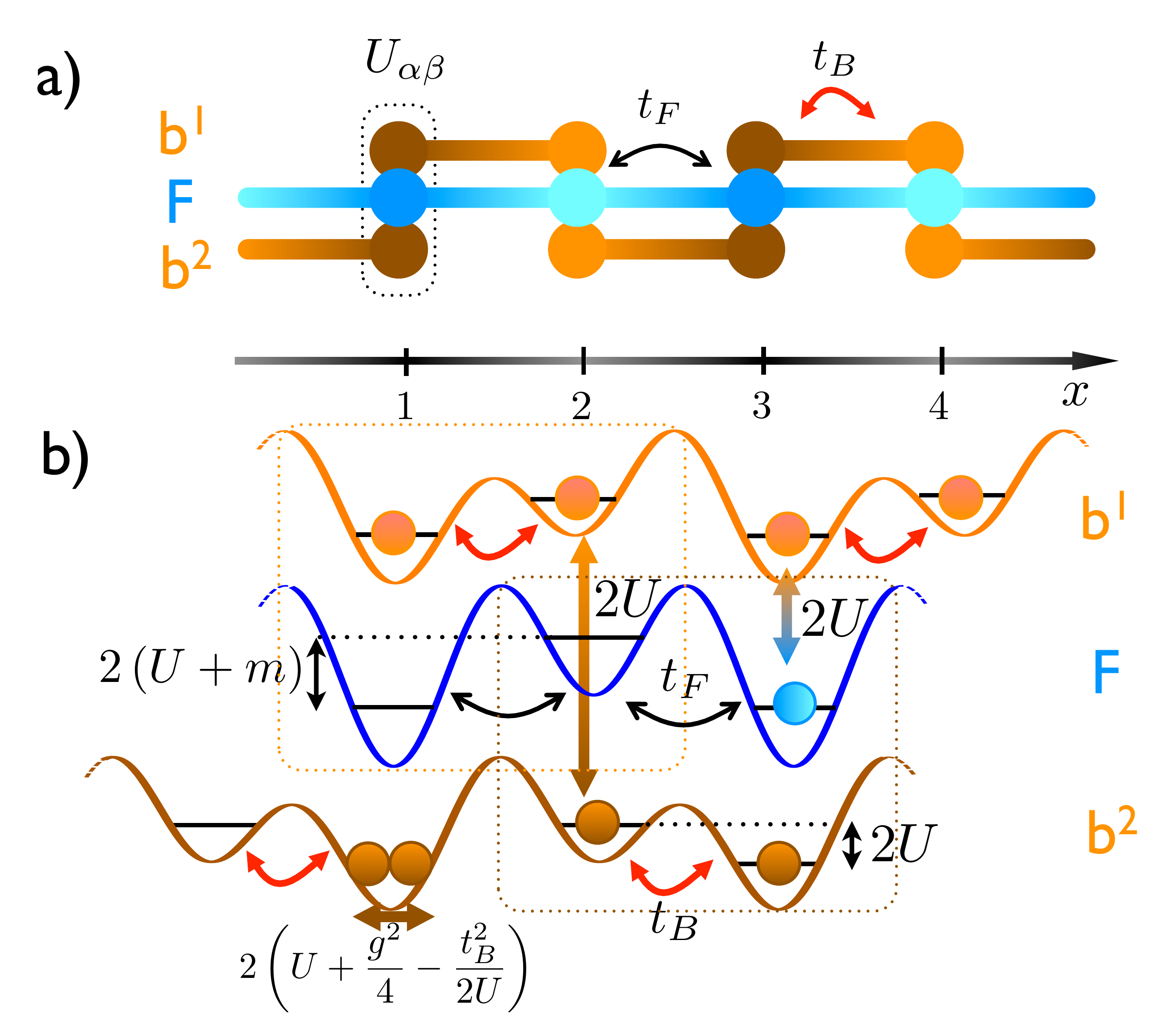}
\caption{[Color online] { Schematic view of the optical super-lattices for one fermionic and two bosonic species 1 and 2 (model I). a) Species 1 can hop between an even site $x$ and the odd site $x-1$, while species 2 can hop between $x$ and $x+1$. b) Illustration of various contributions to the Hamiltonian. Fermions and two-component bosons have on-site repulsions $U_{1F} = U_{2F} = U_{12} = 2U$, while bosons of the same species have $U_{11} = U_{22} = 2U + g^2/2 - t_B^2/U$. The offsets of the bosonic and fermionic super-lattices are $2U_1 = 2U_2 = 2U$ and $2 U_F = 2(U+m)$, respectively. If the fermion hops to the left, it picks up the energy offset $2U$ from a boson of species 2 which simultaneously tunnels to the right.}}
\label{implement}
\end{figure}

The realization of an atomic LGT simulator requires: {\it (i)} the identification of physical degrees of freedom to represent fermionic particles and bosonic quantum link variables; {\it (ii)} to impose the Gauss law in order to remove the gauge variant states; and {\it (iii)} to design the desired dynamics in the gauge invariant subspace. Below we develop a rather general atomic toolbox to implement $U(1)$ lattice gauge models coupled to matter fields based on mixtures of cold fermionic and bosonic atoms in optical lattices. Within this toolbox, we consider two different microscopic realizations in terms of Hubbard models, model I and II. Below we present in some detail the conceptually simpler model I (see Fig. \ref{implement}), which assumes two-component bosons representing gauge fields. Model II, discussed in the \cite{supmat}, assumes one component bosons with magnetic or electric dipolar interactions; it offers better scalability and experimental feasibility. Our concepts generalize immediately to experiments in 2D and 3D, and to fermions with spin~\cite{supmat}.

\begin{figure}[tbp]
\includegraphics[width=0.42\textwidth]{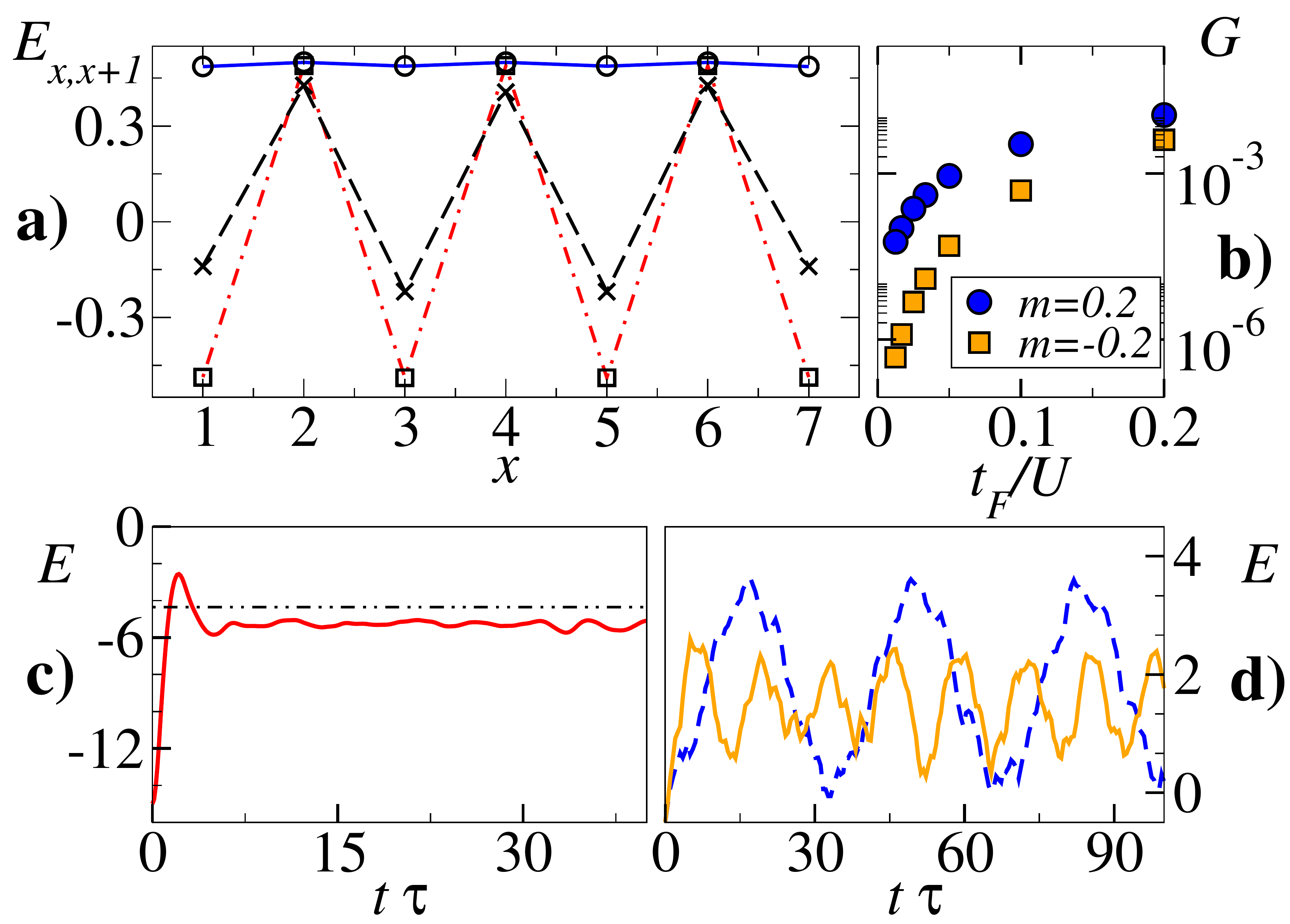}
\caption{[Color online] { {\it a)} Flux configuration in the ground state of Eq.\ref{H_micro} compared to the QLM for $S = \frac{1}{2}$ obtained by exact diagonalization of an $L = 8$ site system. The pa\-ra\-me\-ters of the QLM (in units of $t_F = t_B = 1$) are $t = 0.05, \delta_F = -0.05$ (see \cite{supmat}), and $m = -0.2, 0, 0.2$ (squares, crosses, and circles). The corresponding microscopic pa\-ra\-me\-ters are $U = 20$ and $m= - 0.2, 0, 0.2$ (dashed-dotted, dashed, and solid lines). {\it b)} accuracy of the effective gauge invariance pa\-ra\-me\-ter $G=\sum_x |\langle G_x \rangle|/L$ in the microscopic realization as a function of $t_F/U$.{\it c, d)} Real-time evolution of the total electric flux $E=\sum_x E_{x,x+1}$ obtained by exact diagonalization of the QLM with $L=16$. {\it c)} For $S = 1$ (solid line) string breaking is illustrated, starting from the initial state at the top of Fig.~\ref{cartoon}c, and approaching the corresponding vacuum expectation value (dashed-dotted line) of $E = \sum_x E_{x,x+1}$ ($g^2 = \sqrt{2} t > 0$, $m = 0$, $\delta_{F} = - \sqrt{2} t$; critical breaking length $L_c=0$ when $t=0$). {\it d)} For $S = \frac{1}{2}$ we show the evolution after a quench, starting from the initial state at the top of Fig. \ref{cartoon}d. The flux order pa\-ra\-me\-ter performs coherent oscillations whose period and strength strongly depends on $m$ ($m/t = 0.6 (0.9)$ for dashed (thick) line, $\delta_F = 10 \ t$).}}
\label{results}
\end{figure}

{\it (i)} The spin $S = \frac{1}{2},1,\dots$ representing the quantum link can be realized with a fixed number $N = 2S$ of bosonic atoms in a double well potential with tunnel coupling (Fig. \ref{cartoon}b). An optical super-lattice \cite{Anderlini2007,Trotzky2008} (Fig. \ref{implement}) provides an array of double wells with different depths, and a Mott insulator phase of bosons allows loading with the desired number of atoms $N$. For two neighboring sites $x$ and $x+1$, with $b^\sigma_x$ and $b^\sigma_{x+1}$ denoting the boson destruction operators in the corresponding wells, we define a Schwinger representation for the quantum link
\begin{equation}
\label{bosons}
U_{x,x+1} = b^{\sigma \dagger}_{x+1} b^\sigma_x, \
E_{x,x+1} = \frac{1}{2}\left(b^{\sigma \dagger}_{x+1} b^\sigma_{x+1} - 
b^{\sigma \dagger}_x b^\sigma_x\right).
\end{equation}
The electric flux is related to the population difference of the two sites. Here the bosonic species index $\sigma = 1, 2$ distinguishes between links originating from even and odd sites $x$. Eq.\ (\ref{bosons}) requires that each boson can tunnel only to one specific neighboring site, based on a term $h^B_{x,x+1} = - t_B b^{\sigma \dagger}_{x+1} b^\sigma_x + \mathrm{h.c.}$ The number of bosonic atoms is conserved locally on each link. In \cite{supmat} we discuss model II with just a single bosonic species, by encoding $\sigma$ in the geometric location of the bosons to the left or to the right of the site $x$. We now also add spinless fermionic atoms at half-filling to our super-lattice setup, which can hop between neighboring sites based on the term $h^F_{x,x+1} = - t_F \psi_{x+1}^\dagger \psi_x + \mathrm{h.c}$. {\it (ii)} Gauss law: Using $b^{\sigma \dagger}_x b^\sigma_x + b^{\sigma \dagger}_{x+1} b^\sigma_{x+1} = 2S$, the gauge generator reduces to
\begin{equation}
\widetilde G_x = n_x^F + n^1_x + n^2_x - 2S +
\frac{1}{2}\left[(-1)^x - 1 \right].
\end{equation}
Here $n_x^\alpha$ counts the atoms of type $\alpha = F, 1, 2$. Up to an $x$-dependent constant, $\widetilde G_x$ thus counts the total number of atoms at the site $x$. To impose the Gauss law, we will consider interaction terms which can be rewritten in the form $U \widetilde{G}_x^2$ as the dominant term in the Hamiltonian, so that all gauge variant states are removed from the low-energy sector. This is reminiscent of the repulsive Hubbard model for a Mott insulator \cite{lewenstein2012ultracold}. In this sense, the gauge invariant states (which obey $n_x^F + n^1_x + n^2_x =  2S + \frac{1}{2}\left[1 - (-1)^x \right]$) can be viewed as ``super-Mott'' states. {\it (iii)} It is well known that, for large on-site repulsion, the Hubbard model reduces to the $t$-$J$ model \cite{assa1994interacting}. We now induce the dynamics of a $U(1)$ QLM in a similar manner, by considering the 1D microscopic Hamiltonian $\widetilde H = \sum_x h^B_{x,x+1} + \sum_x h^F_{x,x+1} + m \sum_x (-1)^x n_x^F + U \sum_x \widetilde G_x^2$. Up to an additive constant, it can be expressed as
\begin{eqnarray}
\label{H_micro}
\widetilde H&=&
- t_B \!\! \sum_{x \ \mathrm{odd}} {b^1_x}^\dagger b^1_{x+1}
- t_B \!\! \sum_{x \ \mathrm{even}} {b^2_x}^\dagger b^2_{x+1} 
- t_F \sum_x \! \psi_x^\dagger \psi_{x+1}
\nonumber \\
&+&\mathrm{h.c.} + \sum_{x,\alpha,\beta} n^\alpha_x U_{\alpha \beta} n^\beta_x +
\sum_{x,\alpha }(-1)^x U_\alpha n^\alpha_x.
\end{eqnarray}
The last two terms describe repulsive on-site interactions as well as super-lattice offsets, and form the basic building block for the Gauss constraint $ U \sum_x \widetilde G_x^2$. The various contributions to the Hamiltonian are illustrated in Fig. \ref{implement}b. The QLM of Eq.\ (\ref{Hamilton}) with $t = t_B t_F/U$ emerges in second order perturbation theory, if one tunes the pa\-ra\-me\-ters to the values listed in Fig. \ref{implement}b. The offsets $U_\alpha$ give rise to an alternating super-lattice for both the fermions and the bosons. In analogy to super-exchange interactions \cite{Trotzky2008}, energy conservation enforces a \emph{correlated hop} of the fermion with the spin-flip on the link, thus realizing the term $- t \psi_x^\dagger U_{x,x+1} \psi_{x+1}$. This is the key ingredient for the coupling of fermions and quantum links. Additionally, a gauge invariant term $\delta_{F} \sum_{x} \psi^{\dagger}_{x} \psi_{x} \left[ 1-\psi^{\dagger}_{x+1} \psi_{x+1} \right]$ is also generated~\cite{supmat}.
The reduction of the microscopic model of Eq.\ (\ref{H_micro}) to the QLM of Eq.\ (\ref{Hamilton}) has been verified both at the few- and many-body level, is schematically illustrated in Fig.~\ref{results}a-b and extensively discussed in \cite{supmat}.

We have performed exact diagonalizations on small system sizes to quantitatively show the physical phenomena of string breaking and the dynamics after a quench which can be observed in an experiment. The main results are presented in Fig. \ref{results}c-d. For $S = 1$, we evolve a {\it string state} initially prepared as in Fig. \ref{cartoon}c under Hamiltonian pa\-ra\-me\-ters such that the separation between charge and anti-charge is larger than the characteristic scale for string breaking  $L = 4m/g^2 + 3$. Indeed, the large negative electric flux initially stored in the string quickly approaches its vacuum value, illustrating the string breaking mechanism. For $S = \frac{1}{2}$, Fig. \ref{results}d also shows the time evolution after a quench, starting from the parity-invariant state at the top of Fig. \ref{cartoon}d. In fact, the electric flux, which is an order pa\-ra\-me\-ter for spontaneous parity breaking, displays coherent oscillations, reminiscent of a disoriented chiral condensate in QCD \cite{Rajagopal1993}.
A general experimental implementation, which will require three basic steps (preparation of an initial gauge invariant state, evolution via quantum link dynamics, and measurement of relevant physical observables), is discussed in the supplementary materia\cite{supmat}.

In the present work, we have proposed a quantum simulator of lattice gauge theories, where bosonic gauge fields are coupled to fermionic matter, allowing demonstration experiments for phenomena such as time-dependent string breaking and the dynamics after a quench. While the basic elements behind our model have been demonstrated individually in the laboratory, the combination of these tools and the extension to higher dimensions remain a challenge to be tackled in future generations of optical lattice experiments. While building a QCD quantum simulator to address questions related to non-zero baryon density and real-time evolution remains a long term goal, we see no fundamental obstacles on the atomic physics side, but rather a long list of challenges such as incorporation of multi-component quark fields and non-Abelian plaquette terms in higher dimensions. A realistic pathway will be the investigation of increasingly complex (quantum link) models in an interplay between theory and experiment, with the short term goals of extending the present study to higher dimensions and in particular non-Abelian gauge field models.

{\em Acknowledgment:} We thank  D.\ B.\ Kaplan, M.\ Lewenstein, B.\ Pasquiou, F.\ Schreck, and M.\ Zaccanti for discussions. PZ and MD thank the Joint Quantum Institute for hospitality. Work at Bern is supported by the Schweizerischer Nationalfonds. Work at Innsbruck is supported by the integrated project AQUTE, the Austrian Science Fund through SFB F40 FOQUS, and by the DARPA OLE program. MM is supported by QUITEMAD S2009-ESP-1594, PICC: FP7 2007-2013 (grant Nr.\ 249958) and MICINN grant FIS2009-10061. Authors are listed in alphabetical order. {\em Note added:} While completing the present work, we became aware of two preprints \cite{Zohar2012,Tagliacozzo2012} on atomic quantum simulation of $U(1)$ gauge theories (without coupling to fermions).

\clearpage

\section{Supplementary Information}

\subsection{Symmetries of the U(1) quantum link model}
In this section, we briefly review the basic symmetry properties of the $U(1)$ quantum link model of Eq.\ (3) of the main text. 
\begin{enumerate}
\item As in any gauge theory, the Hamiltonian is invariant against local symmetry transformations. In this case, it commutes with the infinitesimal $U(1)$ gauge generators (additional details are given in the second supplementary material file):
\begin{eqnarray}
&&\widetilde{G}_x = \psi_x^\dagger \psi_x + \frac{1}{2} \left[(-1)^x - 1\right] 
- E_{x,x+1} + E_{x-1,x}.
\end{eqnarray}

\item The parity transformation P is implemented as
\begin{eqnarray}
&&^{P}\psi_x = \psi_{-x},
~~~^{P}\psi_x^{\dag} = \psi_{-x}^{\dag}, \nonumber \\
&&^{P}U_{x,x+1} = U_{-x-1,-x}^{\dag},
~~^{P}E_{x,x+1} = - E_{-x-1,-x},
\end{eqnarray}

\item while charge conjugation C acts as
\begin{eqnarray}
&&^{C}\psi_x = (-1)^{x+1}\psi_{x+1}^{\dag},
~~~^{C}\psi_x^{\dag} = (-1)^{x+1} \psi_{x+1}, \nonumber \\
&&^{C}U_{x,x+1} = U_{x+1,x+2}^{\dag},
~~^{C}E_{x,x+1} = - E_{x+1,x+2}.
\end{eqnarray}

\item For $m = 0$ the Hamiltonian also has a $\Z(2)$ chiral symmetry which shifts all fields by one lattice spacing,
\begin{eqnarray}
&&^{\chi}\psi_x = \psi_{x+1},
~~~^{\chi}\psi_x^{\dag} = \psi_{x+1}^{\dag}, \nonumber \\
&&^{\chi}U_{x,x+1} = U_{x+1,x+2},
~~^{\chi}E_{x,x+1} =  E_{x+1,x+2}.
\end{eqnarray}
However, this symmetry is explicitly broken when one imposes the Gauss law $\widetilde{G}_x|\Psi\rangle = 0$. 

\end{enumerate}

\subsection{Model I: Quantum link model emerging from a Hubbard-type model}

In this section, we sketch the main steps to reduce the \emph{microscopic} Hubbard model, Eq.\ (6) of the main text, to an \emph{effective} quantum link model at low energies using second order perturbation theory. We are interested in the scenario where the largest energy scale $U$ is given by the diagonal Hamiltonian
\begin{eqnarray}
\widetilde{H}_U&=& \!\! \left(\! U \!\! + \! \frac{g^2}{4} \! \right) 
\!\!\! \sum_{x,\sigma=1,2} \!\!\!\! \left(n^\sigma_x \right)^2 \! 
+ \! 2 U \! \sum_x n^1_x n^2_x 
+ U \!\!\!\!\! \sum_{x,\sigma=1,2} \!\!\!\! \left(-1\right)^x \! n^\sigma_x  
\nonumber \\ 
&+&2 \ U \sum_{x,\sigma=1,2} n^F_x n^\sigma_x +
\left(U + m \right) \sum_x \left(-1 \right)^x n^F_x \nonumber \\
&=&\left(U + \frac{g^2}{4}\right) \sum_x \left(E_{x-1,x}^2 + E_{x,x+1}^2\right) 
\nonumber \\
&+&U \sum_x \left(-1 \right)^x \left(E_{x-1,x} - E_{x,x+1}\right) \nonumber \\
&+&2 \ U \sum_x \left[\psi^\dagger_x \psi_x 
\left(E_{x-1,x} - E_{x,x+1}\right) - E_{x-1,x} E_{x,x+1}\right] \nonumber \\
&+&\left(U + m\right) \sum_x \left(-1 \right)^x \psi^\dagger_x \psi_x \nonumber \\
&=&U \sum_x \widetilde{G}_x^2 + \frac{g^2}{2} \! \sum_x E_{x,x+1}^2 + 
m \sum_x \left(-1 \right)^x \psi^\dagger_x \psi_x.
\end{eqnarray}
The values $g^{2}$ and $m$ are small compared to $U > 0$, i.e. $g^2, |m| \ll U$, but they are still relevant in the induced quantum link model.

The term to be generated in second order perturbation theory is the correlated hopping of fermions mediated by the quantum link (represented by a quantum spin). It appears as an effective interaction induced by the previous Hamiltonian and the perturbation terms
\begin{eqnarray}
\Delta \widetilde{H}&=&- t_F \sum_x \left(\psi^\dagger_{x+1} \psi_x + 
\psi^\dagger_x \psi_{x+1} \right) \nonumber \\
&&- t_B \sum_{x \ \text{odd}} \left[b^{1 \dagger}_x b^1_{x+1} + 
b^{1 \dagger}_{x+1} b^1_{x}\right] \nonumber \\
&&- t_B \sum_{x \ \text{even}} \left[b^{2 \dagger}_x b^2_{x+1} + 
b^{2 \dagger}_{x+1} b^2_{x} \right].
\end{eqnarray}
To second order in $t_F$ and $t_B$, the effective Hamiltonian reads
\begin{eqnarray}
H_{\text{eff}}&=& \left( \frac{g^2}{2} + \frac{t_B^2}{U}\right) \sum_x E_{x,x+1}^2 +
m \sum_x \left(-1 \right)^x \psi^\dagger_x \psi_x \nonumber \\
&&\!\!\!\!\!\!\! - \frac{t_F t_B}{U} \sum_x 
\left[\psi_x^\dagger U_{x,x+1} \psi_{x+1} + 
\psi_{x+1}^\dagger U^\dagger_{x,x+1} \psi_x\right] \nonumber \\
&&\!\!\!\!\!\!\! - \frac{t_F^2}{U} \sum_x \psi^\dagger_x \psi_x 
\left(1 - \psi^\dagger_{x+1} \psi_{x+1} \right).
\end{eqnarray}
The last term proportional to $\delta_F = t_F^2/U$ was not present in the original quantum link model Hamiltonian. This is no problem, because this term is also gauge invariant, and could have been added to the quantum link Hamiltonian from the beginning. 

\begin{figure}[tbp]
\includegraphics[width=0.48\textwidth]{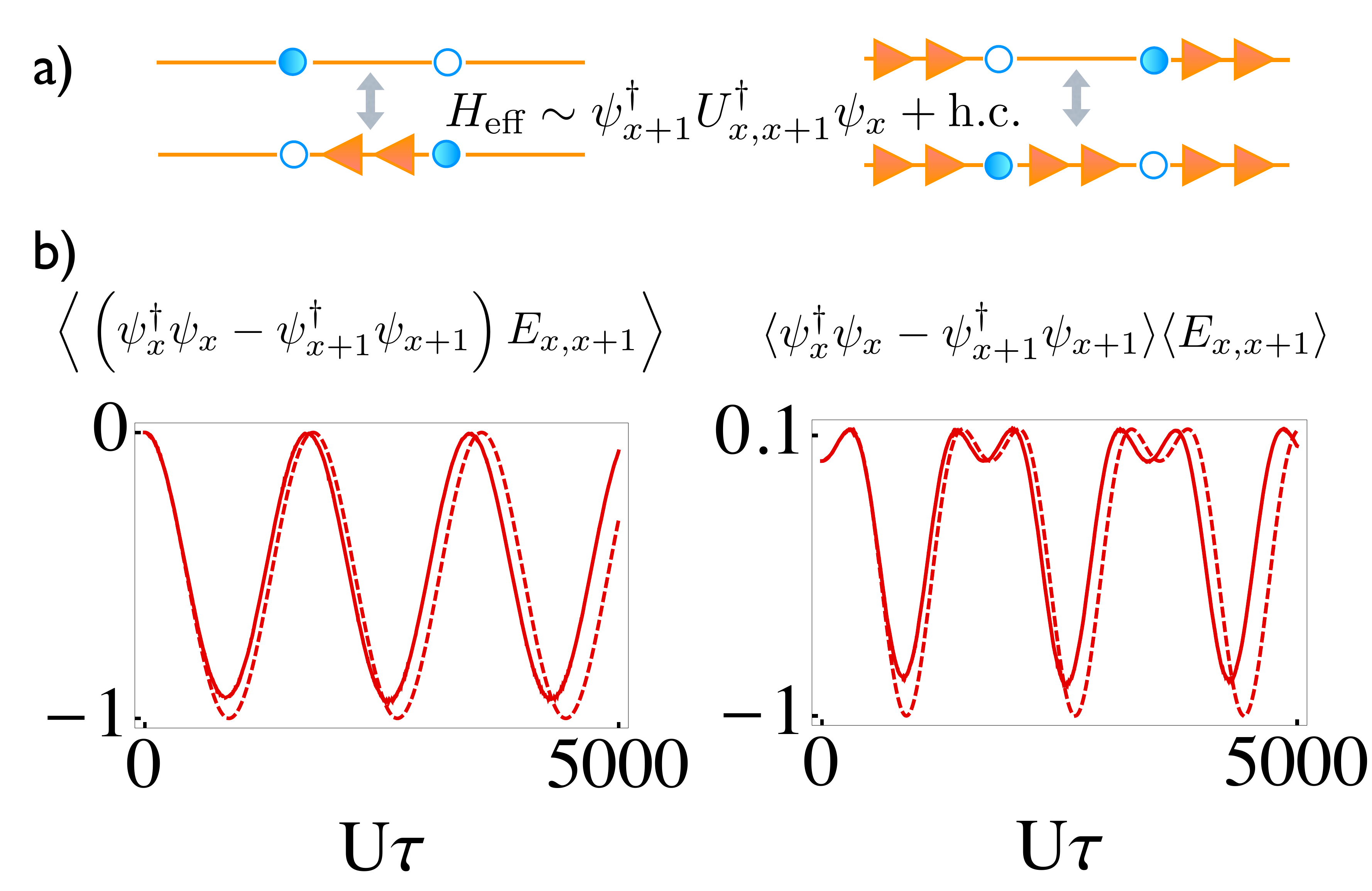}
\caption{[color online] { a) Pictorial representation of four gauge invariant states coupled in second order perturbation theory in the $S = 1$ case. b) Expectation values of the charge difference $\psi^\dagger_x  \psi_x - \psi^\dagger_{x+1} \psi_{x+1} $ and the electric flux $E_{x,x+1}$ as a function of the real-time $\tau$, starting from one of the four states, under the evolution of the microscopic Hamiltonian for $t_F = U/20$ (solid line). The effective quantum link model (dashed line) yields an expectation value $- \frac{1}{2} \left[1 - \cos\left(\sqrt{2} t_F t_B \tau/U\right) \right]$ for the product of both operators (left plot), while the product of both expectation values is $- \frac{1}{2} \cos\left(\sqrt{2} t_F t_B \tau/U\right) \left[\cos\left( \sqrt{2} t_F t_B \tau/U\right) - 1 \right]$ (right plot), signaling the collective dynamics of the coherent fermion hopping mediated by the quantum link. In the plots $t_{F}=2t_{B}$.}}
\label{4state}
\end{figure}

To test the reduction of the microscopic Hubbard-type model to the effective quantum link model, we have considered a minimal setup of four lattice sites in the $S = 1$ case, as illustrated in Fig.\ \ref{4state}a. We have compared the low-energy spectrum of the microscopic and the corresponding effective Hamiltonian. The spectra coincide for $U \gg t_F,t_B$, and even in the intermediate interaction regime $U \simeq 10 \ t_F$ the deviations are of order $1\%$. In view of experimental realizations, another relevant question is to what extent the Hamiltonian preserves the Gauss law. We have considered several initial gauge invariant states, evolving in time according to $\widetilde{H}_U + \Delta \widetilde{H}$. For $U = 10 \ t_F$ ($20 \ t_F$), the probability to leave the gauge invariant subspace is below $10\%$ ($2\%$) even for time scales of order $\tau \simeq 5000~t^{-1}$. Moreover, as demonstrated in Fig\ref{4state}b, the expectation values of $n^F_x$ and $E_{x,x+1}$ display oscillations typical of a coherent two-body process, in direct analogy with the double well experiments in \cite{Anderlini2007s,Trotzky2008s}.

At the many-body level, we have also studied the flux configuration in the ground state of the microscopic Hamiltonian compared to the emergent QLM using exact diagonalization with open boundary conditions in the $S = \frac{1}{2}$ representation. As illustrated in Fig. 3a of the main text, the microscopic model compares very favorably with its QLM analog, and gauge invariance is effectively realized (see Fig. 3b of the main text).

\subsubsection{Implementation procedure}

An experimental implementation will require three basic steps: preparation of an initial gauge invariant state, evolution via quantum link dynamics, and measurement of relevant physical observables. The first step can be implemented by preparation of Mott insulator states of bosonic and fermionic species on different lattice sites via loading in deep optical lattice potentials. Subsequently, the ground state or quench dynamics can be realized via adiabatic or rapid lowering of the depth of the optical lattices. Coherent evolution according to the QLM may be validated at the few-body level by performing double well experiments (corresponding to a single quantum link) along the lines of Ref.~\cite{Anderlini2007s,Trotzky2008s}. A numerical case study is presented in the supplementary information. Finally, {\it in-situ} site-resolved imaging of bosonic particle number distributions \cite{Bakr2009a,Bakr2010,Sherson2010,Weitenberg2011} allows one to measure $E_{x,x+1}$ and to reconstruct the spin-flux configuration and, thus to quantitatively probe the system. 

\subsubsection{Remarks on statistics}
As far as 1D implementations are considered, the statistics of all basic components does not play a fundamental role (this is not true in 2D and 3D, where Fermi statistics is a key-ingredient of matter fields). In particular, the gauge bosons may be substituted by fermions (as far as $S=1/2$ representations are concerned) and matter fields may be represented by hard-core bosons. Such flexibility may be relevant when considering specific experimental implementations, consistently enlarging the set of candidate systems.

\subsection{Model II: Quantum link models in dipolar systems}

Here we illustrate an alternative route toward realizing $U(1)$ quantum link models where a {\em single} dipolar bosonic species is sufficient to realize a gauge covariant link structure. For simplicity, we focus on the $S = \frac{1}{2}$ setup, although larger spins can, in principle, be achieved by considering on-site bosonic interactions. While this construction extends to higher dimensions in a straightforward manner, here we discuss the simpler 1D implementation. 

\begin{figure}[tbp]
\includegraphics[width=0.4\textwidth]{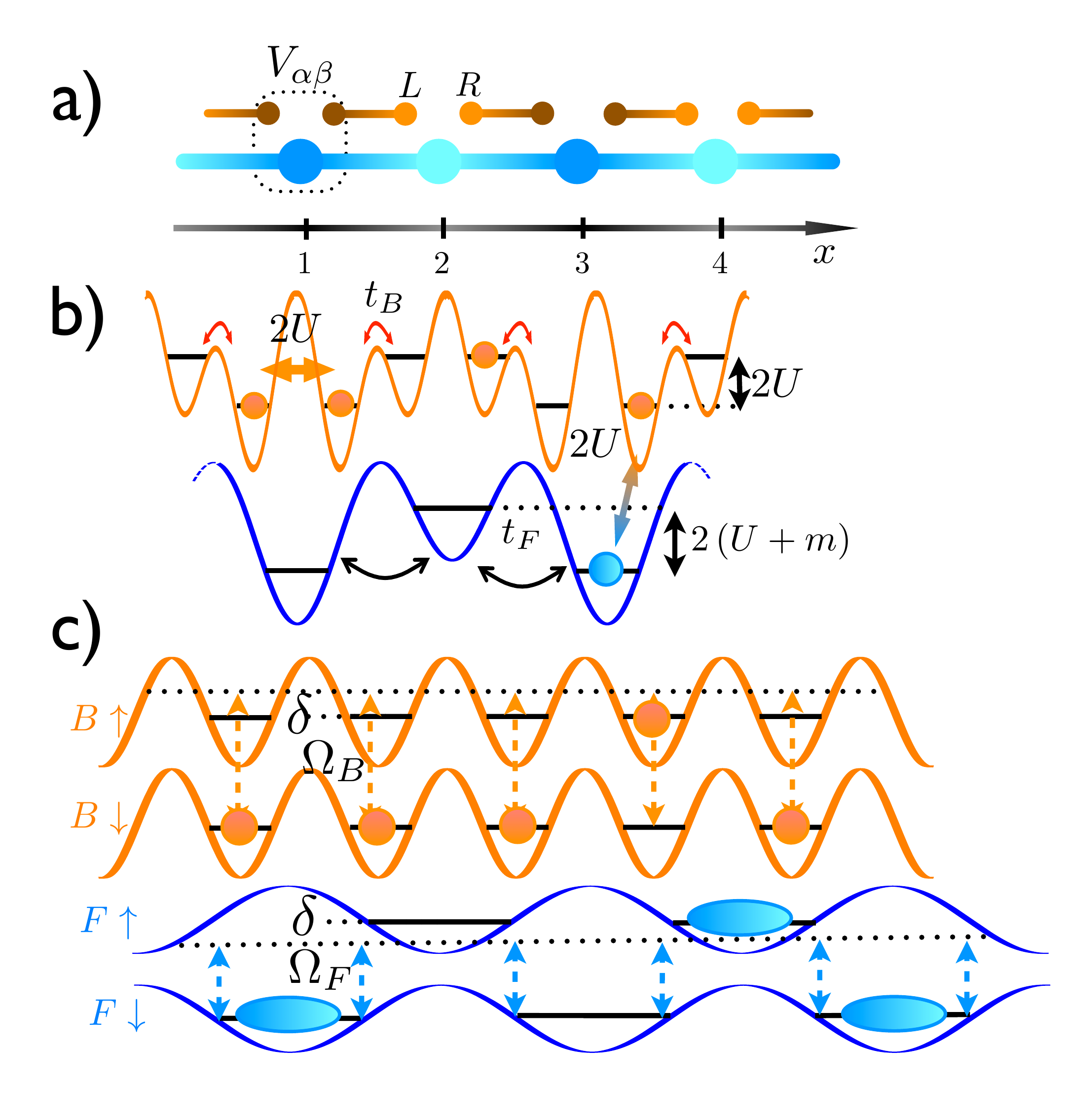}
\caption{[color online] { a) Lattice configuration of model II. Fermions hop between sites $x$ and $x+1$, while bosons hop in double wells on each link. b) Lattice configuration including Hamiltonian parameters. c) Alternative scheme employing Raman assisted tunneling rates: fermions with spin are loaded into a species-dependent optical lattice, while bosons are confined in a periodic potential with half the wavelength.}}
\label{si}
\end{figure}

The microscopic model studied here uses a mixture of fermionic and bosonic particles in the presence of strong dipolar interactions \cite{Lahaye2009s,baranovreview2012s}. Possible experimental realizations are quantum gases of magnetic atoms like Cr \cite{Griesmaier2005s}, Er \cite{Aikawa2012s}, or Dy \cite{Lu2011s, Lu2012s}, and dipolar molecules  \cite{Wang2004s,Sage2005s,Rieger2005s,Deiglmayr2008s,Kraft2006s,VandeMeerakker2008s,Deiglmayr2010s,Lercher2011s,DeMiranda2011s,Chotia2012s}. As sketched in Fig.\ \ref{si}, the mixture is confined to a lattice, where fermions occupy sites labeled by $x$ and bosons are defined on the link sites $(x,L)$ and $(x,R)$, to the left and to the right of $x$. Bosons can hop only between sites $(x,R)$ and $(x+1,L)$, and serve as natural link variables when expressed in terms of Schwinger bosons
\begin{eqnarray}
U_{x,x+1} = b^\dagger_{x+1,L} b_{x,R}, \quad 
E_{x,x+1} = \frac{1}{2}\left(n_{x+1,L} - n_{x,R}\right).
\end{eqnarray}
Note that here the bosonic index $R,L$ is related to the lattice configuration, and, in contrast to Eq.\ (6) of the main text, it is not associated with an internal degree of freedom. By identifying $b_{x,L} = b_x^1$, $b_{x,R} = b_x^2$ (for $x$ even) and $b_{x,L} =b_x^2$, $b_{x,R} = b_x^1$ (for $x$ odd), we can relate the bosons with spatial indices $(x,L)$ and $(x,R)$ to the bosonic species $\sigma = 1,2$ that arise in model I discussed in the main text. The microscopic Hamiltonian of model II takes the form
\begin{eqnarray}
\label{H_dip}
\widetilde{H}_{\textrm{dip}}&=&- t_F\sum_{x}(\psi^\dagger_x\psi_{x+1} + 
\textrm{h.c.}) + m \sum_x (-1)^x n_x^F \nonumber \\
&&- t_B\sum_x (b_{x,L}^\dagger b_{x+1,R} + \textrm{h.c.}) + 
\!\!\!\! \sum_{x,\alpha=L,R} 
\!\!\!\! \omega_\alpha(-1)^x n_{x,\alpha} \nonumber \\
&&+ 2 U \sum_{x \leq y} \sum_{\alpha,\beta=L,R} 
n_{x,\alpha} v_{\alpha\beta}[x,y] n_{y,\beta} \nonumber \\
&&+ 2W_{FB} \sum_x n_x^F [n_{x,L} + n_{x,R}],
\end{eqnarray}
where $v_{\alpha\beta}[x,y]$ is given by the dipolar interaction between the particles. Its strength is normalized such that $v_{RL}[x,x+1] = 2 v_{LR}[x,x] = 1$. The Bose-Fermi interaction is a combined effect of both dipolar and short-range potentials. The latter stems from the partial overlap between the single-site Wannier functions of bosons and fermions in $(x,L,R)$ and $x$, respectively. In analogy with model I, and taking into consideration the fast spatial decay of dipolar interactions, one can reformulate the Hamiltonian of Eq.\ (\ref{H_dip}) as $\widetilde{H}_{\textrm{dip}}= U \sum_x \widetilde{G}_x^2 + \Delta \widetilde{H}_{\textrm{dip}}$, where 
\begin{equation}
\widetilde{G}_x = n_x^F + n_{x,L} + n_{x,R} + \frac{1}{2}\left[(-1)^x + 1\right].
\end{equation}
By choosing $\omega_F = U + m$, $W_{FB} =\omega_{L,R} = U$, one then obtains a quantum link model with $t = 2t_B t_F/U$, with additional gauge invariant diagonal terms generated by the dipolar interaction beyond nearest-neighbor sites.

In case of {\em magnetic} atoms \cite{Lu2011s,Aikawa2012s,Griesmaier2005s}, the interaction regime $U \gtrsim 10 \ t_F$ may be achieved by properly tuning the interspecies scattering length and the optical lattice depth, leading to typical energy scales of the order $5\:\textrm{nK}$ for, e.g., Dy bosonic gases confined in an optical lattice with a lattice spacing of about $200 \:\textrm{nm}$ \cite{Lu2012}. Polar molecules have large {\em electric} dipole moments which can be aligned by using external electric fields providing sufficiently strong constraint energies $U$ \cite{baranovreview2012s} when loaded into optical lattices \cite{Chotia2012s}. A clear advantage of this setup is that it can be straightforwardly adapted to 2D, since, in contrast to model I, just one bosonic species is required regardless of the dimensionality. Furthermore, dipolar interactions are isotropic once the dipole moments are aligned perpendicular to the 2D plane \cite{Lahaye2009s,baranovreview2012s}.

The above realization of gauge fields is based on bosonic atoms in a double well potential corresponding to the familiar {\em external} Josephson effect. Another intriguing possibility provided by dipolar interactions makes use of an {\em internal} Josephson effect \cite{leggett2006quantums}, where (instead of two wells coupled by a tunnel coupling) one considers atoms with two internal states coupled by a Rabi frequency. Such an internal Josephson junction provides an alternative realization of quantum link spins. The setup that we propose is illustrated in Fig.\ \ref{si}. The gauge invariant dynamics are realized by combining a fermionic species {\em with spin} moving in a {\em spin-dependent} optical lattice, and a bosonic dipolar species confined in a deep optical lattice of half the wavelength. Tunneling of fermions between adjacent lattice sites is implemented by a Raman assisted transition in the spirit of \cite{jaksch2003creations,Aidelsburger2011s}, coupled to the {\em internal} Josephson system with the corresponding bosonic link site (c.f.\ Fig.\ \ref{si}), leading to an effective Hamiltonian term of the form $\psi^\dagger_x U_{x,x+1} \psi_{x+1}$. Gauge invariance is then implemented by considering state-dependent dipole-dipole interactions between the bosonic links, and a finite detuning shift for the Raman transition $\delta$. A detailed study of this implementation scheme will be reported elsewhere.

\clearpage

\section{String Breaking and False Vacuum Decay in the $(1+1)$-d $U(1)$ Quantum Link Model with Staggered Fermions}

\subsection{Introduction}

String breaking is an important dynamical mechanism in Quantum Chromodynamics (QCD). A quark and an anti-quark separated by a distance $r$ are connected by a string of color electric flux. The string costs energy in proportion to its length, and thus leads to a linearly rising quark-anti-quark potential $V(r) = \sigma r$, where $\sigma$ is the string tension. When the quark mass is taken to infinity, i.e.\ when one considers an $SU(3)$ Yang-Mills theory of pure gluons, in which quarks and anti-quarks appear just as external sources of color flux, the confining string is unbreakable. In QCD with dynamical quarks, however, the string can break by quark-anti-quark pair creation, as soon as the energy of the string is larger than the mass of the mesons generated in the pair creation process. The dynamics of string breaking, i.e.\ its evolution in real time, is inaccessible to classical simulations using lattice gauge theory, which is usually formulated in Euclidean time. As for other quantum systems with a large number of degrees of freedom, studying the real-time evolution of a lattice gauge theory leads to a very severe sign problem. For this reason, understanding the real-time dynamics of gauge theories from first principles remains inaccessible to the traditional methods of theoretical physics. Quantum simulators are a very promising tool for overcoming this problem. String
breaking is only one example of a dynamical mechanism whose real-time evolution is interesting but very difficult to investigate. The real-time evolution of heavy ion collisions is another example, which may serve as a strong motivation and an ultimate long-term goal for the development of quantum simulators for gauge theories.

Another interesting phenomenon of great relevance in particle physics is the spontaneous breaking and restoration of symmetries. For example, in the early universe, chiral symmetry was intact before it got spontaneously broken at sufficiently low temperature. In heavy ion collisions, one attempts to recreate a droplet of the early unbroken phase and observes its evolution back into the vacuum state. The time evolution of metastable false vacuum states has been discussed in doomsday scenarios of cosmic evolution, in which the rapid expansion of a bubble of true vacuum wipes out the preexisting false vacuum. In one spatial dimension, string breaking by pair creation and false vacuum
decay are similar phenomena, with the dynamical quark and anti-quark playing the role of the walls of the true vacuum bubble. In fact, as we will see, we can use the same model Hamiltonians to quantum simulate both string breaking and false vacuum decay. 

In Wilson's formulation of lattice gauge theory, the fundamental gauge degrees of freedom are parallel transporter matrices taking values in the gauge group, which are associated with the links connecting neighboring lattice sites on a 4-dimensional Euclidean space-time lattice. In lattice QCD the link variables are $3 \times 3$ unitary matrices with determinant 1, which constitute an 8-parameter family of continuously varying classical gauge degrees of freedom per link. As a consequence, in Wilson's formulation of lattice gauge theory, there is an infinite-dimensional Hilbert space already for each individual link, which largely complicates the construction of quantum simulators.

Quantum link models are an alternative formulation of lattice gauge theory, in which the fundamental gauge degrees of freedom are represented by discrete quantum variables, so-called quantum links, which have a finite-dimensional Hilbert space per link. In the simplest case of a $U(1)$ gauge theory, a quantum link is simply represented by a quantum spin. When one chooses spin $\frac{1}{2}$, a single qubit per link is hence sufficient to represent the gauge field. In an $SU(2)$ gauge theory the quantum link is a generalized quantum spin with a continuous gauge symmetry, which has at least 4 discrete degrees of freedom. Hence, two qubits per link are required to represent a  non-Abelian gauge field. A $U(3)$ quantum link model requires 6 and an $SU(3)$ quantum link model requires 20 discrete states per link, which amounts to 3 or 5 qubits per link, respectively. For simplicity, in this supplementary material we limit ourselves to $U(1)$ quantum link models in one spatial dimension, considering both the spin $\frac{1}{2}$ representation (one qubit) and the spin 1 representation, with 3 discrete states per link. 

In this supplementary material, we discuss quantum link models with a small number of degrees of freedom, which still display the physical phenomenon of string breaking or false vacuum decay. The proposed model Hamiltonians are not yet fully realistic from a particle physics point of view. In particular, they address the physics of the Schwinger model (i.e.\ QED in $1+1$ space-time dimensions) rather than QCD in $3+1$ dimensions. For simplicity, the Schwinger model is formulated with so-called staggered fermions, which have only one degree of freedom per lattice site. It should be pointed out that we are not yet addressing the Schwinger model in the continuum limit. Hence, several steps will have to be taken in order to turn the proposed quantum simulations into something that becomes truly useful for particle physics. Still, the pathway towards this ultimate goal is clearly visible. 

\subsection{From Classical Background Gauge Fields to Quantum Links}

In order to better understand the theoretical framework of quantum link models, let us begin to discuss the familiar situation of fermions moving in a classical background magnetic field described by a vector potential $\vec A$ with $\vec B = \vec \nabla \times \vec A$. For simplicity, we consider spinless fermions hopping on a lattice with sites $x$. The fermion creation and annihilation operators obey standard anti-commutation relations 
\begin{equation}
\{\psi_x,\psi_y^\dagger\} = \delta_{xy}, \quad 
\{\psi_x,\psi_y\} = \{\psi_x^\dagger,\psi_y^\dagger\} = 0.
\end{equation}
The term in the Hamiltonian that describes gauge covariant hopping from a site $y$ to a nearest-neighbor site $x$ is then given by $\psi_x^\dagger u_{xy} \psi_y$. Here $u_{xy} = \exp(i \varphi_{xy}) \in U(1)$ is a classical gauge parallel transporter associated with the link connecting $x$ and $y$ that is obtained by integrating the vector potential along the link
\begin{equation}
\varphi_{xy} = \int_x^y d\vec l \cdot \vec A.
\end{equation}
After a gauge transformation $\vec A' = \vec A - \vec \nabla \alpha$ one obtains
\begin{equation}
\varphi_{xy}' = \int_x^y d\vec l \cdot (\vec A - \vec \nabla \alpha) =
\varphi_{xy} + \alpha_x - \alpha_y,
\end{equation} 
and hence the parallel transporter transforms as
\begin{equation}
u_{xy}' = \exp(i \varphi_{xy}') = \exp(i \alpha_x) u_{xy} \exp(- i \alpha_y).
\end{equation}
In order to render the hopping Hamiltonian gauge invariant, the gauge transformation of the classical background field must be accompanied by the gauge transformation of the fermion creation and annihilation operators
\begin{equation}
\psi_x' = \exp(i \alpha_x) \psi_x, \quad 
{\psi_x^{\dagger}}' = \psi_x^\dagger \exp(- i \alpha_x),
\end{equation}
which leaves their anti-commutation relations unchanged.

In this work, we are not limiting ourselves to classical background fields. Instead the electromagnetic field is treated as a dynamical entity with its own quantum dynamics that is intimately connected to the dynamics of the fermions. A dynamical lattice gauge field $U_{xy}$ is no longer given in terms of a classical background field $\vec A$. Instead, the link variable $U_{xy}$ is an independent quantum degree of freedom, whose canonically conjugate ``momentum'' variable $E_{xy}$ represents an electric field operator that obeys the commutation relation
\begin{equation}
\label{commutator}
[E_{xy},U_{xy}] = U_{xy}.
\end{equation}
In a gauge theory one uses redundant gauge variant variables to describe the gauge invariant physics. In order to eliminate the unphysical gauge variant states, one must impose Gauss' law. Physical states $|\Psi\rangle$ are gauge invariant, i.e.\ $G_x |\Psi\rangle = 0$, where
\begin{equation}
G_x = \psi_x^\dagger \psi_x - \sum_i \left(E_{x,x+\hat i} - E_{x-\hat i,x}\right)
\end{equation}
is the infinitesimal generator of gauge transformations. A general (non-infinitesimal) gauge transformation with parameters $\alpha_x$ is described by the unitary transformation
\begin{equation}
V = \prod_x \exp(i \alpha_x G_x),
\end{equation}
which indeed acts as
\begin{eqnarray}
\label{gaugetrafo}
U_{xy}^\prime &&= V^\dagger U_{xy} V = \exp(i \alpha_x) U_{xy} \exp(-i \alpha_y), 
\nonumber \\
\psi_x^\prime &&= V^\dagger \psi_x V = \exp(i \alpha_x) \psi_x, \nonumber \\
{\psi_x^\dagger}^\prime &&= V^\dagger \psi_x^\dagger V = \psi_x^\dagger \exp(i \alpha_x).
\end{eqnarray}

The Hamiltonian of a lattice gauge theory with a dynamical $U(1)$ gauge field coupled to dynamical fermions takes the form
\begin{eqnarray}
\label{Ham}
H =&& - t \sum_{\langle x y \rangle} 
\left(\psi_x^\dagger U_{xy} \psi_y + \mathrm{h.c.}\right) +
\frac{g^2}{2} \sum_{\langle x y \rangle} E_{xy}^2 \nonumber \\
&&- \frac{1}{4 g^2} \sum_{\langle w x y z \rangle} \left(U_{wx} U_{xy} U_{yz} U_{zw} + \mathrm{h.c.}\right).
\end{eqnarray}
Here $t$ is the hopping amplitude and $g$ is the gauge coupling that determines the electric and the magnetic field energies. By construction, the Hamiltonian is gauge invariant, i.e.\ $[H,G_x] = 0$.

In the Hamiltonian formulation of Wilson's $U(1)$ lattice gauge theory, the link variables are still complex phases $U_{xy} = \exp(i \varphi_{xy}) \in U(1)$, but the link angles $\varphi_{xy}$ are now independent degrees of freedom (unrelated to a background field $\vec A$). Correspondingly, the canonically conjugate electric field operators are given by $E_{xy} = - i \partial/\partial \varphi_{xy}$, such that the commutation relation of eq.(\ref{commutator}) is indeed satisfied. Since Wilson's parallel transporter is a continuous variable, the corresponding Hilbert space is infinite-dimensional even for just a single link. In the quantum link formulation of $U(1)$ lattice gauge theory, on the other hand, the dimension of the link Hilbert space is finite, and given by the $2S+1$ states of an integer or half-integer quantum spin $\vec S_{xy}$ on each link. In this case, the electric field operator is given by $E_{xy} = S_{xy}^3$ with eigenvalues $-S,\dots,S$, while the gauge field is represented by the quantum link operators
\begin{equation}
U_{xy} = S_{xy}^+ = S_{xy}^1 + i S_{xy}^2, \quad
U_{xy}^\dagger = S_{xy}^- = S_{xy}^1 - i S_{xy}^2,
\end{equation}
which act as raising and lowering operators of electric flux. It is to be noted that the operators $E_{xy}, U_{xy}, U^{\dagger}_{xy}$ satisfy a local SU(2) algebra at each link (see Eq. \ref{commutator}). This allows the realization of a local U(1) gauge invariance of the Hamiltonian in Eq. \ref{Ham}.

\subsection{Lattice Hamiltonian for the Schwinger Model with Staggered Fermions}

The Hamiltonian for a simple quantum link model is given by
\begin{eqnarray}
H =&& -t \sum_{x} \left[ \psi_x^{\dag} U_{x,x+1} \psi_{x+1} + 
\psi_{x+1}^{\dag}U_{x,x+1}^{\dag} \psi_x\right] \nonumber \\
&&+ m \sum_x (-1)^x \psi_x^{\dag}\psi_x + \frac{g^2}{2}\sum_{x} E_{x,x+1}^2.
\end{eqnarray}
This model describes a single species of staggered fermions (i.e., the fermion field operator $\psi$ has only a single component) minimally coupled to a $U(1)$ gauge field. The electric flux operator defined on each link is $E_{x,x+1} = S_{x,x+1}^3$, and the gauge field is represented by quantum links $U_{x,x+1} = S_{x,x+1}^+$ and $U_{x,x+1}^{\dag} = S_{x,x+1}^-$, which satisfy the commutation relation
\begin{equation}
[E_{x,x+1},U_{y,y+1}]= \delta_{xy} U_{x,x+1}.
\end{equation}

In this supplementary material, we will consider quantum links both in the spin $\frac{1}{2}$ and in the spin 1 representation. In the minimal spin $\frac{1}{2}$ representation, the allowed electric flux eigenvalues are $\pm \frac{1}{2}$, i.e.\ it is impossible to have zero flux. This may seem unnatural from the point of view of particle physics. Still, as we will see, the resulting physics resembles that of a non-zero vacuum angle $\theta$. The spin 1 representation allows the three flux eigenvalues 0 and $\pm 1$.

The factor $(-1)^x$ in the fermion mass term is due to the use of staggered fermions. It explicitly breaks the translation symmetry by one lattice spacing. This symmetry plays the role of a discrete $\Z(2)$ chiral symmetry for staggered fermions. While in the spin 1 case the dynamics of the gauge field is governed by the term proportional to the gauge coupling $g^2$, this term has a trivial contribution in the spin $\frac{1}{2}$ case.

\subsection{Symmetries of the Hamiltonian}

The following symmetry transformations are interesting to consider:

\begin{itemize}

\item Gauge Invariance: The Hamiltonian commutes with the infinitesimal generators of local $U(1)$ transformations
\begin{equation}
G_x = \psi_x^\dag \psi_x - E_{x,x+1} + E_{x-1,x}.
\end{equation}

\item Chiral Symmetry: The model admits the following discrete chiral symmetry transformation:
\begin{eqnarray}
&&^{\chi}\psi_x = \psi_{x+1},
~~~^{\chi}\psi_x^{\dag} = \psi_{x+1}^{\dag},  \nonumber \\ 
&&^{\chi}U_{x,x+1} = U_{x+1,x+2},
~~^{\chi} E _{x,x+1} = E_{x+1,x+2}.
\end{eqnarray}
This transformation preserves all terms in the Hamiltonian except the mass term.

\item Charge conjugation C: This is again a discrete symmetry, which is implemented as:
\begin{eqnarray}
&&^{C}\psi_x = (-1)^{x+1}\psi_{x+1}^{\dag},
~~~^{C}\psi_x^{\dag} = (-1)^{x+1}\psi_{x+1}, \nonumber \\
&&^{C}U_{x,x+1} = U_{x+1,x+2}^{\dag},
~~^{C}E_{x,x+1} = - E_{x+1,x+2}.
 \end{eqnarray}
This transformation leaves all terms invariant.

\item Parity P: The parity transformation is implemented as:
\begin{eqnarray}
&&^{P}\psi_x = \psi_{-x},
~~~^{P}\psi_x^{\dag} = \psi_{-x}^{\dag}, \nonumber \\
&&^{P}U_{x,x+1} = U_{-x-1,-x}^{\dag},
~~^{P}E_{x,x+1} = - E_{-x-1,-x}.
\end{eqnarray}

\item CP Symmetry: The combined symmetry takes the form:
\begin{eqnarray}
&&^{CP}\psi_x = (-1)^{-x+1} \psi_{-x+1}^{\dag},
~~~^{CP}\psi_x^{\dag} = (-1)^{-x+1} \psi_{-x+1}, \nonumber \\
&&^{CP}U_{x,x+1} = U_{-x,-x+1},
~~^{CP}E_{x,x+1} = E_{-x,-x+1}.
\end{eqnarray}

\end{itemize}

\subsection{Implementation of Gauss' Law}

\begin{figure}[h]
\centering

\setlength{\unitlength}{3pt}
\begin{picture}(30,25)(0,0)

\multiput(0,5)(0,10){2}{\line(1,0){10}}
\put(4.0,5){\thicklines \color{blue}\vector(-1,0){0}}
\put(6.5,15){\thicklines \color{blue}\vector(1,0){0}}

\multiput(20,0)(0,10){3}{\line(1,0){10}}
\put(23.0,0){\thicklines \color{blue}\vector(-1,0){0}}
\put(25.0,0){\thicklines \color{blue}\vector(-1,0){0}}
\put(25.5,20){\thicklines \color{blue}\vector(1,0){0}}
\put(27.5,20){\thicklines \color{blue}\vector(1,0){0}}

\end{picture}

\caption{Left: Spin $\frac{1}{2}$ quantum link carrying an electric flux of 
$+ \frac{1}{2}$ or $- \frac{1}{2}$,
respectively. Right: Spin 1 quantum link with a flux of +1, 0, or -1, 
respectively.}
\label{fig1}
\end{figure}
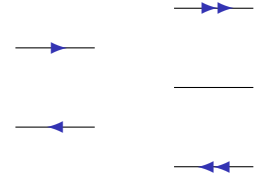

To set the conventions, we will use the pictorial representation for the quantum links shown in Figure \ref{fig1}.

Before we discuss the implementation of the Gauss law, note that the staggered fermions introduce a ``staggering'' in the occupation numbers of the vacuum, which resembles a half-filled state in condensed matter physics, even in the absence of the gauge field. Because of the mass term, in the vacuum (i.e.\ the Dirac sea) the fermions occupy the odd sites (for $m, t > 0$, $m \gg t$). This means that the Gauss law cannot be satisfied in the usual way: even with equal amounts of electric flux leaving and entering a site, the flux divergence $G_x = \psi_x^\dag \psi_x - E_{x,x+1} + E_{x-1,x}$ depends on whether the site is even or odd. This is an unavoidable feature of the staggered fermion formulation. We define the Gauss law such that the configuration with the staggered occupation in the absence of electric flux is a physical state that satisfies it, although not in the usual form. All physical states $|\Psi\rangle$ are then required to satisfy
\begin{eqnarray}
&&\widetilde G_x |\Psi\rangle = \nonumber \\
&&\left[\psi_x^\dag \psi_x - E_{x,x+1} + E_{x-1,x} + \frac{ (-1)^x - 1}{2}\right]
|\Psi\rangle = 0.
\end{eqnarray}
The vacuum state for spin 1 and $t=0$, which obeys Gauss' law, is illustrated in Figure \ref{fig4}a.

\subsection{String Breaking}

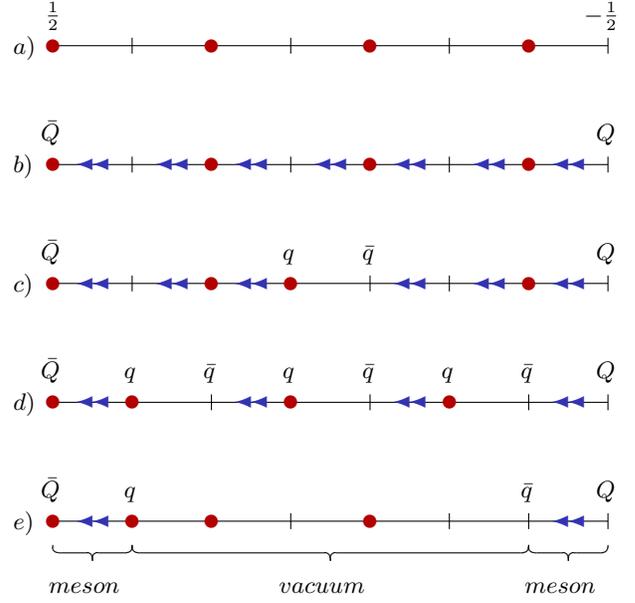
\begin{figure}
\centering
\setlength{\unitlength}{3pt}
\begin{picture}(80,75)(-5,-10)

\multiput(0,0)(0,15){5}{\line(1,0){70}}

\multiput(0,60)(20,0){4}{\color{red}\circle*{1.5}}
\multiput(10,60)(20,0){4}{\put(0,-1){\line(0,1){2}}}
\put(-5,59){$a)$}
\put(-1,63){$\frac{1}{2}$}
\put(67,63){$-\frac{1}{2}$}

\multiput(0,45)(20,0){4}{\color{red}\circle*{1.5}}
\multiput(10,45)(20,0){4}{\put(0,-1){\line(0,1){2}}}
\multiput(3.0,45)(10,0){7}{\thicklines \color{blue}\vector(-1,0){0}}
\multiput(5.0,45)(10,0){7}{\thicklines \color{blue}\vector(-1,0){0}}
\put(-5,44){$b)$}
\put(-1.5,48){$\bar Q$}
\put(68.5,48){$Q$}

\multiput(0,30)(20,0){2}{\color{red}\circle*{1.5}}
\put(30,30){\color{red}\circle*{1.5}}
\put(60,30){\color{red}\circle*{1.5}}
\put(10,30){\put(0,-1){\line(0,1){2}}}
\put(40,30){\put(0,-1){\line(0,1){2}}}
\multiput(50,30)(20,0){2}{\put(0,-1){\line(0,1){2}}}
\multiput(3.0,30)(10,0){3}{\thicklines \color{blue}\vector(-1,0){0}}
\multiput(5.0,30)(10,0){3}{\thicklines \color{blue}\vector(-1,0){0}}
\multiput(43.0,30)(10,0){3}{\thicklines \color{blue}\vector(-1,0){0}}
\multiput(45.0,30)(10,0){3}{\thicklines \color{blue}\vector(-1,0){0}}
\put(-5,29){$c)$}
\put(-1.5,33){$\bar Q$}
\put(29,33){$q$}
\put(39,33){$\bar q$}
\put(68.5,33){$Q$}

\put(0,15){\color{red}\circle*{1.5}}
\multiput(10,15)(20,0){3}{\color{red}\circle*{1.5}}
\multiput(20,15)(20,0){3}{\put(0,-1){\line(0,1){2}}}
\put(70,15){\put(0,-1){\line(0,1){2}}}
\multiput(3.0,15)(20,0){4}{\thicklines \color{blue}\vector(-1,0){0}}
\multiput(5.0,15)(20,0){4}{\thicklines \color{blue}\vector(-1,0){0}}
\put(-5,14){$d)$}
\put(-1.5,18){$\bar Q$}
\put(9,18){$q$}
\put(19,18){$\bar q$}
\put(29,18){$q$}
\put(39,18){$\bar q$}
\put(49,18){$q$}
\put(59,18){$\bar q$}
\put(68.5,18){$Q$}

\multiput(0,0)(20,0){3}{\color{red}\circle*{1.5}}
\put(10,0){\color{red}\circle*{1.5}}
\multiput(30,0)(20,0){3}{\put(0,-1){\line(0,1){2}}}
\put(60,0){\put(0,-1){\line(0,1){2}}}
\put(3.0,0){\thicklines \color{blue}\vector(-1,0){0}}
\put(5.0,0){\thicklines \color{blue}\vector(-1,0){0}}
\put(63.0,0){\thicklines \color{blue}\vector(-1,0){0}}
\put(65.0,0){\thicklines \color{blue}\vector(-1,0){0}}

\put(-5,-1){$e)$}
\put(-1.5,3){$\bar Q$}
\put(9,3){$q$}
\put(59,3){$\bar q$}
\put(68.5,3){$Q$}

\put(1,-3){\oval(2,2)[bl]}
\put(1,-4){\line(1,0){3}}
\put(4,-5){\oval(2,2)[tr]}
\put(6,-5){\oval(2,2)[tl]}
\put(6,-4){\line(1,0){3}}
\put(9,-3){\oval(2,2)[br]}
\put(-0.5,-9){$meson$}

\put(11,-3){\oval(2,2)[bl]}
\put(11,-4){\line(1,0){23}}
\put(34,-5){\oval(2,2)[tr]}
\put(36,-5){\oval(2,2)[tl]}
\put(36,-4){\line(1,0){23}}
\put(59,-3){\oval(2,2)[br]}
\put(28.5,-9){$vacuum$}

\put(61,-3){\oval(2,2)[bl]}
\put(61,-4){\line(1,0){3}}
\put(64,-5){\oval(2,2)[tr]}
\put(66,-5){\oval(2,2)[tl]}
\put(66,-4){\line(1,0){3}}
\put(69,-3){\oval(2,2)[br]}
\put(59.5,-9){$meson$}

\end{picture}
\caption{\it a) vacuum, b) string induced by a static external $\bar Q Q$ pair, c) broken string, d) evolution, e) final state with two mesons separated by vacuum.}
\label{fig4}
\end{figure}

The string formed between an external static charge-anti-charge pair is an interesting physical object to be studied using quantum simulation. Let us consider string breaking in the case of spin 1 quantum links. Figure \ref{fig4}b shows an initial state which has a string joining an external static charge $Q$ and an anti-charge $\overline Q$ placed at both ends. As the fermions become lighter, the sequence of figures \ref{fig4}c -- \ref{fig4}e shows the breaking of the string due to the creation of fermion-anti-fermion pairs, and the evolution to the final state of two mesons at the ends. For the static $t = 0$ case, the vacuum energy is given by $E_0 = - m L/2$, while the energies of the initial unbroken string state and the final two-meson state take the form
\begin{eqnarray}
E_{\text{string}} - E_0 &&= \frac{g^2}{2} (L - 1), \nonumber \\
E_{\text{mesons}} - E_0 &&= 2 (\frac{g^2}{2} + m).
\end{eqnarray}
The critical distance for string breaking is determined by the condition $E_{\text{string}} - E_{\text{mesons}} = \frac{g^2}{2} (L - 3) - 2m$ and thus results in $L = 4m/g^2 + 3$.

\subsection{False Vacuum Decay in a Minimal Model with C and P Symmetry}

The minimal model with $S = \frac{1}{2}$ quantum links may seem unnatural from a particle physics point of view, because it does not allow zero flux. However, it resembles a non-zero vacuum angle $\theta = \pi$. In order to explore these dynamics, let us consider the Hamiltonian
\begin{eqnarray}
H =&& -t \sum_{x} \left[ \psi_x^{\dag} U_{x,x+1} \psi_{x+1} + 
\psi_{x+1}^{\dag}U_{x,x+1}^{\dag} \psi_x\right]  \nonumber \\
&&+  m \sum_x (-1)^x \psi_x^{\dag}\psi_x - 2 \sigma \sum_x (-1)^x E_{x,x+1}.
\end{eqnarray}
The last term favors a staggered flux pattern and is charge conjugation and parity invariant. It does, however, explicitly break chiral symmetry. Since chiral symmetry is anyways explicitly broken by the Gauss law implementation, this additional breaking is unimportant. Terms linear in the electric field are characteristic of a non-zero vacuum angle $\theta$. For $\theta \neq 0, \pi$, charge conjugation and parity are both explicitly violated. At $\theta = 0, \pi$, on the other hand, charge conjugation and parity remain exact symmetries, as it is the case here. Interestingly, in the Schwinger model at $\theta = \pi$, both charge conjugation and parity are spontaneously broken. As we will see, this is also the case in the model that we have just introduced, at least for sufficiently large values of the mass $m$ (and for $t \neq 0$), while for small $m$, charge conjugation and parity are restored. Hence, the model with the staggered flux term proportional to $\sigma$ mimics the physics of $\theta = \pi$, which is again very difficult to simulate with a classical computer, due to a very severe sign problem. Let us finally notice that, after implementing Gauss law, the staggered electric field term can be recast into a renormalization for the fermionic mass term, that is $m_{\textrm{eff}}=m-\sigma$, so that one recovers exactly the model in Eq. (3) of the main text. 

\begin{figure}[h]
\centering
\setlength{\unitlength}{3pt}
\begin{picture}(80,6)(0,-5)
\put(0,0){\line(1,0){80}}
\multiput(0,0)(20,0){5}{\color{red}\circle*{1.5}}
\multiput(10,0)(20,0){4}{\put(0,-1){\line(0,1){2}}}
\multiput(6.5,0)(20,0){4}{\thicklines \color{blue}\vector(1,0){0}}
\multiput(14.0,0)(20,0){4}{\thicklines \color{blue}\vector(-1,0){0}}
\put(-1,-5){$L$}
\put(9,-5){$1$}
\put(19,-5){$2$}
\put(29,-5){$...$}
\put(65,-5){$L-1$}
\put(79,-5){$L$}
\end{picture}
\caption{C and P invariant candidate vacuum state of the spin $\frac{1}{2}$ model.}
\label{fig6}
\end{figure}
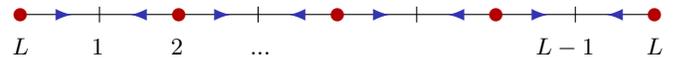

\begin{figure}
\centering
\setlength{\unitlength}{3pt}
\begin{picture}(85,21)(-5,-5)

\multiput(0,0)(0,15){2}{\line(1,0){80}}

\multiput(0,15)(20,0){5}{\put(0,-1){\line(0,1){2}}}
\multiput(10,15)(20,0){4}{\color{red}\circle*{1.5}}
\multiput(4.0,15)(10,0){8}{\thicklines \color{blue}\vector(-1,0){0}}
\put(-5,14){$a)$}
\put(-1,10){$L$}
\put(9,10){$1$}
\put(19,10){$2$}
\put(29,10){$...$}
\put(65,10){$L-1$}
\put(79,10){$L$}

\multiput(0,0)(20,0){5}{\put(0,-1){\line(0,1){2}}}
\multiput(10,0)(20,0){4}{\color{red}\circle*{1.5}}
\multiput(6.5,0)(10,0){8}{\thicklines \color{blue}\vector(1,0){0}}
\put(-5,-1){$b)$}
\put(-1,-5){$L$}
\put(9,-5){$1$}
\put(19,-5){$2$}
\put(29,-5){$...$}
\put(65,-5){$L-1$}
\put(79,-5){$L$}

\end{picture}
\caption{Competing vacua of the spin $\frac{1}{2}$ model, which are C and P partners of each other.}
\label{fig7}
\end{figure}
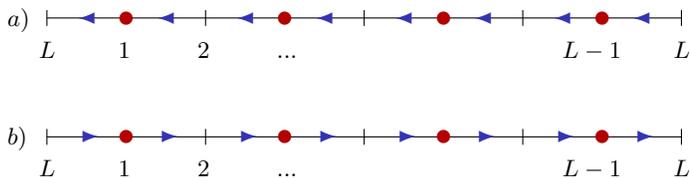

Let us consider the physics in the vacuum sector, using periodic boundary conditions of a finite system with an even number of lattice sites $L$. For $t = 0$, there are three candidate vacuum states, which are depicted in Figures \ref{fig6}
as well as \ref{fig7}a and \ref{fig7}b. 
First of all, it is instructive to convince oneself that these three states indeed satisfy the Gauss law. The state shown in Figure \ref{fig6} is invariant under both charge conjugation and parity, while the two states in Figure \ref{fig7} are charge conjugation and parity partners of each other. At $t = 0$ the energies of the three states are given by
\begin{equation}
E_0 = - \sigma L + m \frac{L}{2}, \quad 
E_0' = E_0'' = - m \frac{L}{2},
\end{equation}
such that $E_0' - E_0 = E_0'' - E_0 = (\sigma - m) L$. Hence, for $m < \sigma$ (with $\sigma > 0$) the state depicted in Figure \ref{fig6} is the true vacuum (i.e.\ the Hamiltonian eigenstate of lowest energy respecting the Gauss law). In this case, both C and P are unbroken. On the other hand, for $m > \sigma$, the two states shown in Figure \ref{fig7} have a lower energy and thus C and P are then spontaneously broken. This is reminiscent of the situation in the Schwinger model at $\theta = \pi$. It would be interesting to quantum simulate the decay of the false vacuum of Figure \ref{fig6} into the true vacua of Figure \ref{fig7} in real time. The quench dynamics discussed in the main text is directly related to this phenomenon.

\subsection{Conclusions}

As we have seen, a simple (1+1)-d quantum link model with staggered fermions allows us to address interesting dynamical questions using quantum simulation. Along the way towards addressing similar problems in QCD, one faces the challenges of higher dimensions, non-Abelian gauge fields, and multi-component fermions. In the quantum link formulation of QCD all these elements are present, and we see no fundamental obstacle against implementing then with ultra cold matter in optical lattices. As one goes beyond (1+1) dimensions, gauge field acquire transverse degrees of freedom and display a richer dynamics. Since quantum link models are gauge invariant by construction, these dynamics are those of QCD in the continuum limit. How to best realize this in quantum simulations will be addressed in forthcoming publications.

\end{document}